\documentclass[twocolumn,preprintnumbers,amsmath,amssymb]{revtex4}

\usepackage{graphicx}
\usepackage{dcolumn}
\usepackage{bm}
\def\be{\begin{equation}}
\def\ee{\end{equation}}
\def\bea{\begin{eqnarray}}
\def\eea{\end{eqnarray}}

\begin{document}


\title{The ratio of $\gamma / \pi^0$  production rates in neutrino-nucleus interactions at the $\Delta$ resonance mass region }

\author{Ara N. Ioannisian}

\affiliation{ 
CERN, 
Geneva 23, CH-1211, Switzerland\\
Yerevan Physics Institute, 
Yerevan-36, Armenia\\
 Institute for Theoretical Physics and Modeling, Yerevan-36, Armenia }

\begin{abstract}
We study the dependence of neutrino-induced $\gamma/\pi^0$ production 
($\stackrel{ {(-)}}{\nu_\mu} + A \to \stackrel{ {(-)}}{\nu_\mu}(\mu) +\gamma/\pi^0  + X$) on the target nucleus A,  at the $\Delta$ resonance mass region. 
Our conclusion is based on experimental data for $\pi^0$ production rates at photon-nucleus interactions from the A2 collaboration at the Mainz MAMI accelerator. 
We assume that $\Delta$ resonance decays are independent of the production mechanism (via photon, Z, or W boson). 
In Neutral Current (NC) interactions, the $1\pi^0 + X$ production scales as A$^{2/3}$. 
In contrast, photons from $\Delta$ decays typically escape the nucleus, resulting in a cross-section proportional to the atomic number A.
Thus, in NC interactions, the ratio of $\gamma$ production to $\pi^0$ production is proportional to A$^{1/3}$. 
In Charged Current (CC) $\nu_\mu$($\bar \nu_\mu$) -induced production of $\Delta^+$ ($\Delta^0$) will be proportional to the number of neutrons (protons) in the nucleus. 
After $\Delta$ decay, due to the charge universality in strong interactions, the suppression factor for $\pi^0$ escaping the nucleus must also follow A$^{-1/3}$, as in NC interactions.  \\
We predict the ratio of the  $\gamma / \pi^0$  production rates in NC and CC interactions and for $\nu_\mu$ and $\bar \nu_\mu$ beams:
\begin{description}
\item \hspace {2cm}  {\bf Argon target}: $\sim$3.1\% (NC/CC $\nu_\mu/\bar \nu_\mu$)
\item \hspace {2cm}  {\bf Water target}:  $\sim$1.9\% (NC), $\sim$2.3\% (CC $\nu_\mu$), $\sim$1.7\% (CC $\bar \nu_\mu$)
\item \hspace {2cm}  {\bf Liquid Scintillator target}: $\sim$1.7\% (NC), $\sim$2.1\% (CC $\nu_\mu$), $\sim$1.6\% (CC $\bar \nu_\mu$)
\end{description}
The accuracy of our predictions is limited due to uncertainties in the decay rate $\Delta^{+/0} \to p/n+\gamma$ and the presence of other (minor) channels of neutrino-induced $\gamma / \pi^0$ production in the $\Delta$ mass region. \\
We also discuss solving the MiniBooNE anomaly by looking at the CC single photon and single neutral pion production rates at the SBN program experiments at Fermilab.
\\
\end{abstract}

\maketitle
In neutrino-nucleus interactions, the $\Delta$ resonance plays a crucial role at 4-momentum transfers close to the $\Delta$ resonance mass region.
The T2K (and future HyperK \cite{Hyper-Kamiokande:2018ofw}) neutrino beam has an energy range centered on 600 MeV. The Short Baseline Neutrino (SBN) program at FNAL (SBND, MicroBooNE, and ICARUS) \cite{SBN} is located on the Booster Neutrino Beam at Fermilab, which peaks at 700MeV neutrino energy. DUNE \cite{DUNE:2015lol} experiment's 2nd oscillation maximum also lies at the resonance region. 
The decay products of the $\Delta$ resonance - single photons and neutral pions - constitute significant backgrounds that can lead to misinterpretation of neutrino oscillation results.

\cite{ParticleDataGroup:2022pth} estimated the branching ratio of $\Delta$
to $\gamma$ (from the measured amplitude of the $\gamma +p \to \Delta^+
$) to be
\be
 Br(\Delta^{+/0} \to p/n +\gamma)=(6 \pm 0.5) \cdot 10^{-3}
\ee
while the corresponding branching ratio to $\pi^0$ is
\be
Br(\Delta^{+/0} \to p/n +\pi^0)\simeq 2/3.
\ee
Thus, the ratio of $\Delta^{+/0}$ decaying channels to $\gamma$ or to $\pi^0$ is \cite{Ioannisian:2019kse}
\be
R^0 \equiv {\Gamma^\gamma (\Delta^{+/0})\over \Gamma^{\pi^0}(\Delta^{+/0})} \simeq (9 \pm 0.75) \cdot 10^{-3}.
\label{rr}
\ee
The A2 collaboration \cite{A2} at the Mainz MAMI accelerator investigates photon absorption on nucleons and nuclei. The incident photon beam has a very well-known energy, flux, and polarization in the energy range 40-1603 MeV, yielding precise measurements of the
photon absorption cross sections with different final states.

\begin{figure}[t!]
 \includegraphics[width=0.5\textwidth]{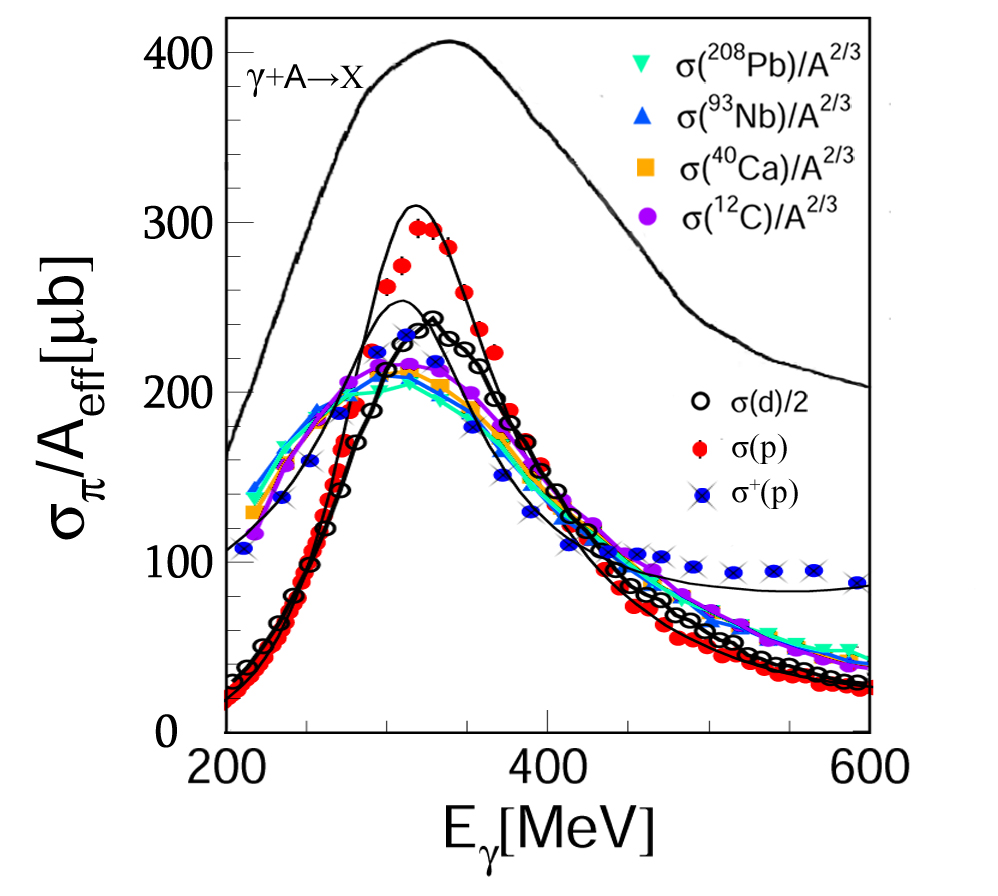}
\caption[...]{All plots are taken from \cite{Krusche:2004xz}. 
The effective mass number $A_{eff}$ is 1 for the proton, 2 for the deuteron, and $A^{2/3}$ for heavier nuclei.  $\sigma^+$ is the cross section of $\gamma +p \to n + \pi^+$.  $\gamma + A\to X$  is (per nucleon) photo-absorption cross section on nuclei, $A_{eff}$=A. 
\label{Fig0}}
\end{figure}

According to the A2 collaboration (Fig 1) \cite{Krusche:2004xz} the $\pi^0$ photo-production on nuclei scales as A$^{2/3}$
\be
\sigma(\gamma + A \to 1\pi^0 +X ) \propto A^{2/3}.
\ee

The photon absorption cross-section on the target nucleus is proportional to the atomic number of the nucleus, A (Fig.1). 
This is an indication of the dominance of incoherent processes. 
The $A^{2/3}$ scaling of the  $\pi^0$ rates (Fig. 1) indicates the suppression of neutral pions escaping the nucleus as an $A^{-1/3}$ factor.
 
 Theoretical calculations \cite{Zhang:2012xi}
show that incoherent $\gamma/\pi^0$ neutrino production processes are dominating over coherent ones at the $\Delta$ resonance mass region. Thus, $\gamma$ production rate must be proportional to the atomic number of the target nucleus, A. Regarding the $\pi^0$ production rate, an additional  $\pi^0$ escape suppression factor must be considered. Based on the results of the A2 experiment \cite{Krusche:2004xz}, in this paper, we will assume the $\pi^0$ escape suppression factor as equal to $ A^{-1/3}$.  We conclude that the ratio of photon
to neutral pion production via a $\Delta$ resonance in nuclei should be
proportional to $A^{1/3}$. 

In the nucleus, nucleons (also $\Delta$s) have slightly different properties than free ones \cite{Zhang:2012xi}. In the current paper, we will neglect this and work on approximating the nucleus as a sum of free nucleons in incoherent interactions.

The $\Delta^{+/0}$ resonances can be produced via Neutral Current (NC) and Charged Current (CC) interactions on the target's protons and neutrons 
  \bea
   p+Z \to \Delta^+ 
 &&  n+Z \to \Delta^0
 \\
   p+W^- \to \Delta^0
 &&  n+W^+ \to \Delta^+
\eea

We define $R$  as a ratio of the $\gamma$ to $\pi^0$  production rates by the (anti)neutrino beam on different targets.

For the CC neutrino beam, we get
\be
\hspace{-0.35cm}R^\mathtt{target}_{CC}\equiv { \sigma_{(\nu+\mathtt{target} \to \mu^-+\gamma +X)} \over \sigma_{ ( \nu + \mathtt{target} \to \mu^-+\pi^0 + X)}}={\sum{N^n_i}\over\sum {N^n_i \cdot A_i^{-1/3}}}R^0 \ ,
\ee
and for the CC anti-neutrino beam, we get
\be
\hspace{-0.35cm}\bar R^\mathtt{target}\equiv { \sigma_{(\bar \nu+\mathtt{target} \to \mu^++\gamma +X)} \over \sigma_{ ( \bar \nu + \mathtt{target} \to \mu^++\pi^0 + X)}} = {\sum{N^p_i}\over\sum {N^p_i \cdot A_i^{-1/3}}}R^0 \ ,
\ee
in the NC, the ratio of the $\gamma/\pi^0$ production rates is equal for neutrino and antineutrino beams
\be
R^\mathtt{target}_{NC}\equiv { \sigma_{(\nu+\mathtt{target} \to \gamma +X)} \over \sigma_{ ( \nu + \mathtt{target} \to \pi^0 +X)}} ={\sum{A_i}\over\sum {A_i^{2/3}}}R^0\ .
\ee
 Here, $A_i$ is the atomic number of the nucleus, $A_i=N^p_i+N^n_i$, $N^p_i$  is the number of protons in the nucleus, 
$N^n_i$  is the number of neutrons in the nucleus.

For water target, ($\mathtt{H_2O}$), we get
\bea
R_{ {\mathtt{NC}}}^{\mathtt{H_2O}}&=&{16+2 \over 16 \cdot 16^{-1/3}+2 }R^0=2.16 R^0=1.9\%
\label{wnc}
\\
R_{ {\mathtt{CC}}}^{\mathtt{H_2O}}&=&{8 \over 8\cdot 16^{-1/3} }R^0=2.52 R^0=2.3\%
\\ 
\bar R_{ {\mathtt{CC}}}^{\mathtt{H_2O}}&=&{8+2 \over 8\cdot 16^{-1/3}+2 }R^0=1.93 R^0=1.7\%
\eea
For liquid scintillator target, (methylene $\mathtt{CH-CH_3}$, in average $\mathtt{C H_2}$), we get
\bea
R_{ {\mathtt{NC}}}^{\mathtt{C H_2}}&=&{12+2 \over 12 \cdot 12^{-1/3}+2 }R^0=1.93 R^0=1.7\%
\label{mini}
\\
R_{ {\mathtt{CC}}}^{\mathtt{C H_2}}&=&{6 \over 6 \cdot 12^{-1/3} }R^0=2.29 R^0=2.1\%
\\
\bar R_{ {\mathtt{CC}}}^{\mathtt{C H_2}}&=&{6+2 \over 6 \cdot 12^{-1/3}+2 }R^0=1.73 R^0=1.6\%
\eea
For the argon target, in all cases (NC or CC interactions, for neutrino or antineutrino beams), we get
\be
\hspace{-0.2cm}
R^{\mathtt{Ar}}=R_{\mathtt{NC}}^{\mathtt{Ar}}=R_{\mathtt{CC}}^{\mathtt{Ar}}=\bar R_{\mathtt{CC}}^{\mathtt{Ar}}=40^{1\over 3}R^0=3.42R^0=3.1\%
\label{arg}
\ee
We have used the central value of (\ref{rr}), $R_0=0.9\%$, for numerical results in (\ref{wnc})-(\ref{arg}). 

The MiniBooNE experiment estimates the number of photons from the registered number of neutral pions. The photons are a background, and an understimation of their possible number could lead to the misinterpretation of the experimental results.  The MiniBooNE experiment used the value $R_{ {\mathtt{NC}}}^{\mathtt{C H_2}}=0.9\%$ \cite{MiniBooNE:2020pnu}.  

The MicroBooNE claims (based on NC interactions) they rejected (94.8\% C.L.) $\Delta \to N \gamma$ as the only source for the MiniBooNE low-energy excess events \cite{MicroBooNE:2021zai}. We estimate that there must be about twice ($1.7/0.9\simeq 1.9$) as many photons in the MiniBooNE as the MiniBooNE estimation. These "additional" photons will not explain all the low-energy excess events, but will substantially lower the number of unexplained electron-like events, moving the "anomaly" to just $\sim$2$\sigma$ statistical fluctuations of the low-energy excess events.

Several theoretical calculations of the $\gamma$ and $\pi^0$ rates at the MiniBooNE experiment exist \cite{Hill:2010zy}. We recomend estimating $R_{ {\mathtt{NC}}}^{\mathtt{C H_2}}$ based on experimentally measured   $R^{\mathtt{Ar}}$. 
The experimentally extracted value of $R^{\mathtt{Ar}}$ (from (\ref{arg})) at the ongoing SBN program experiments \cite{SBN} will allow us to experimentally find the value $R^{exp}_0$.

We want to underline that although there are difficulties in measuring the single photon rate in NC interactions (due to significant systematic errors/background \cite{MicroBooNE:2021zai}), the same ratio, $R^{\mathtt{Ar}}$, can be easily extracted in CC interactions due to the appearance of the muon track from the same vertex. Thus, we recommend that SBN program experiments look at CC single photon and single neutral pion production rates, $R_{\mathtt{CC}}^{\mathtt{Ar}}$ and $\bar R_{\mathtt{CC}}^{\mathtt{Ar}}$. The experimentally extracted value $R^{exp}_0$ can be used to estimate  $R_{ {\mathtt{NC}}}^{\mathtt{C H_2}}$ (\ref{mini}) and shall answer to the long-standing low-energy events excess anomaly at the MiniBooNE. 

Regarding the accuracy of our estimation, we note that additional contributions exist for both  $\gamma$ and $\pi^0$ productions. Neutrino-induced production of neutral pions and photons can be described as resulting from coherent and incoherent processes. The final states of these two processes (including particle content and their momentum spectra) are very different, so there is no interference between them. Coherent processes exist for both $\gamma$ and $\pi^0$ production, but they are subdominant in comparison to the incoherent processes \cite{Zhang:2012xi}.
On the other hand, there are interferences between different channels of incoherent $\gamma/\pi^0$ neutrino-production ($\Delta$ resonance, non-resonance, higher resonances...). But again, as theoretical calculations show, the main contribution is from the $\Delta$ resonance.
Thus, we may conclude that our results/estimations must be highly accurate.


\hspace{0.5cm}

In this note, we discussed the $\Delta$ resonance contribution to the $\gamma$ to $\pi^0$ production rates ratio in neutrino-nucleus interaction. 
We provide semi-quantitative predictions for this ratio in common detector materials such as water, liquid scintillator, and argon. 
A better understanding and measurement of the $\gamma/\pi^0$ production ratio will help constrain background models. They may provide a new avenue for investigating anomalies such as the low-energy events excess anomaly at the MiniBooNE. .


We would like to express our gratitude to Ornella Palamara for the discussions.

This work received partial funding from the Armenian grant 21T-1C333, and we acknowledge COST Action CA23130 support.



\end{document}